\documentclass[prl,twocolumn,showpacs,amsmath,amssymb]{revtex4}

\usepackage{graphicx}
\usepackage{dcolumn}
\usepackage{bm}

\begin{document}

\newcommand{\bn}{{\bf n}}
\newcommand{\bp}{{\bf p}}   
\newcommand{\br}{{\bf r}}
\newcommand{\bk}{{\bf k}}
\newcommand{\bv}{{\bf v}}
\newcommand{\brho}{{\bm{\rho}}}
\newcommand{\wk}{\omega_{\bf k}}
\newcommand{\nk}{n_{\bf k}}
\newcommand{\eps}{\varepsilon}
\newcommand{\la}{\langle}
\newcommand{\ra}{\rangle}
\newcommand{\be}{\begin{eqnarray}}
\newcommand{\ee}{\end{eqnarray}}
\newcommand{\intl}{\int\limits_{-\infty}^{\infty}}
\newcommand{\dE}{\delta{\cal E}^{ext}}
\newcommand{\SE}{S_{\cal E}^{ext}}
\newcommand{\dsp}{\displaystyle}
\newcommand{\phit}{\varphi_{\tau}}
\newcommand{\EF}{{\cal E}_F}
\newcommand{\fn}{f^{(0)}}

\title{ Effects of Electron-Electron Scattering 
        in Wide Ballistic Microcontacts }

\author{ K. E. Nagaev$^1$ and O. S. Ayvazyan$^2$ }
\affiliation{$^1$Institute of Radioengineering and Electronics,  Mokhovaya 11-7, Moscow, 125009 Russia\\
             $^2$Moscow Institute of Physics and Technology,
             Dolgoprudny,  141700 Russia}

\date{\today}

\begin{abstract}
We calculate the corrections to the Sharvin conductance of ballistic multimode microcontacts that result from electron--electron scattering in the leads. Using a semiclassical Boltzmann equation, we obtain that these corrections are {\it positive} and scale with temperature as $T^2\ln(E_F/T)$ for three-dimensional contacts and as $T$ for two-dimensional ones. These results are relevant to recent experiments on 2DEG contacts.
\end{abstract}

\pacs{73.21.Hb, 73.23.-b, 73.50.Lw}

\maketitle

It is a textbook knowledge that normal electron--electron scattering does not contribute to the resistivity of homogeneous metal. However if the metal contains microscopic inhomogeneities, a quantum-mechanical interference between electron--electron and electron--impurity scattering results in a temperature-dependent correction to the conductivity \cite{Altshuler}. The electron--electron scattering is also known to affect the resistance of narrow channels with boundary scattering because it deflects electrons moving along the channel axis to the boundaries, where they can dissipate their momentum \cite{Molenkamp}. Here we address the problem of electron--electron scattering in a confined geometry without any impurity scattering. Namely, we consider a ballistic metal contact with a large number of transverse quantum channels and smooth boundaries. Though collisions between electrons do not change their total momentum, they change their trajectories and hence may prevent some of them from passing through the contact or help some extra electrons to get through. Surprisingly, to the best of our knowledge this problem was never analyzed before.

In the Landauer--Buttiker formalism  the dissipation of power in a contact with a perfect transmission is due to relaxation processes in the leads \cite{Imry}. This relaxation brings the injected electrons in equilibrium with those in the electrodes. It is implicitly taken into account by assuming that electrons incident on the contact from the electrodes have an equilibrium distribution. However this implicit 
relaxation  does not take into account the back action of injected electrons upon  the electrons in the leads, and this is precisely the subject of our paper.

Studies of ballistic microcontacts were pioneered by Sharvin \cite{Sharvin}.
Generally, these contacts were three-di\-men\-si\-onal (3D) and semiclassical, but both experimentalists \cite{Yanson} and theorists \cite{Kulik} focused on the electron--phonon scattering at 
high voltages and ignored the electron--electron interaction, which was less important in this range.

In 1990s of the last century, it became possible to fabricate contacts out of two-dimensional (2D) electron gas. The size of these contacts was comparable to the Fermi wavelength of electrons, and their conductance exhibited steps of size $e^2/h$ as the width increased and new quantum channels opened \cite{steps}. So the main subject of studies became the linear conductance as a function of this width. The effects of electron--electron interaction were considered for such systems mainly in connection with the "0.7 anomaly" \cite{0.7} in terms of a quasi-localized state in the limit of few transverse quantum channels. There was also a great deal of theoretical work on very long and narrow single-channel contacts with electron--electron interactions within the concept of Luttinger liquid \cite{Luttinger}. Very recently, Lunde et al. \cite{Lunde} studied effects of electron--electron scattering in a two-channel long and narrow quantum contact using a kinetic equation with a collision integral. However no calculations in the limit of large channel number have been performed so far. 

\begin{figure}[t]
 \includegraphics[width=8.5cm]{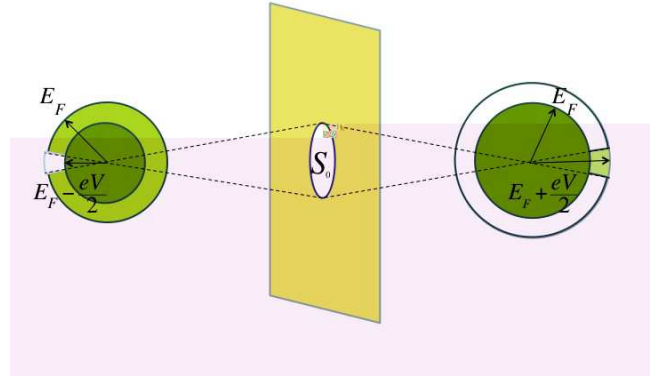}
 \caption{\label{fig1} The distributions of non-interacting electrons to the left and to the right of the orifice  
 \cite{Kulik}.
 In the left half-space, the Fermi sphere has a dip because the electrons injected from the contact have lower  
 energies.  In the right half-space, the Fermi sphere has a bump because the electrons injected from the contact have  
 higher  energies.
 }
\end{figure}

We use an approach similar to that of Kulik {\it et al.} \cite{Kulik} for the case of electron--phonon scattering.
As a model of the contact, we consider two metal half-spaces separated by a thin insulating layer with an orifice. 
We  assume the reflection from the insulator to be specular, but all our results are valid also for diffuse boundary scattering.
The radius $a$ of the orifice is much larger than the Fermi wavelength but much smaller than both elastic and inelastic mean free path of electrons. 
The distribution functions of electrons on both sides of the insulator obey Boltzmann equation
\be
 \frac{\partial f}{\partial t}
 +
 \bv\,\frac{\partial f}{\partial\br}
 +
 e{\bf E}\,\frac{\partial f}{\partial\bp}
 =
 \hat{I}_{ee},
 \label{Boltz}
\ee
where ${\bf E} = -\nabla\varphi$ is the electric field. The electron--electron
collision integral is given by
\begin{eqnarray}
 \hat{I}_{ee}(\bp)
 =
 \alpha_{ee}\,\nu_d^{-2}
 \int\frac{d^dk}{(2\pi)^d}
 \int\frac{d^dp'}{(2\pi)^d}
 \int d^dk'\nonumber\\
 \times
 \delta(\bp + \bk - \bp' - \bk')\,
 \delta( \eps_{\bp} + \eps_{\bk} - \eps_{\bp'} - \eps_{\bk'} )
\nonumber\\
 \times
 \Bigl\{
  [1 - f(\bp)]\,[1 - f(\bk)]\, f(\bp')\, f(\bk')\,
\nonumber\\
  -
  f(\bp)\, f(\bk)\, [1 - f(\bp')]\, [1 - f(\bk')]
 \Bigr\},
 \label{I_ee}
\end{eqnarray}
where $\alpha_{ee}$ is the dimensionless parameter of electron--electron scattering, $d=2$ or 3 is the dimensionality of the system, and $\nu_d$ is the corresponding Fermi density of states. Equation (\ref{Boltz}) should be solved together with the Poisson equation for the electric potential $\varphi$. As the distance from the orifice increases, the distribution function tends to its equilibrium value $f_0(\eps_{\bp})$ and $\varphi$ tends to $eV/2$  and  $-eV/2$ in the left and right half-spaces, $V$ being the voltage drop across the contact. The current through the contact is given by
\be
 I=
 e\int_{S_0} d^{d-1}\rho\,
 \int\frac{d^dp}{(2\pi)^d}\,
 v_{\perp}\, f(\bp,\bm{\rho}),
 \label{current}
\ee
where $p_{\perp}$ is the component of $\bp$ normal to the insulator and vector $\bm{\rho}$ labels points within the orifice area $S_0$.

Because of the condition $E_F \gg{\rm max}(eV, T)$ one may treat the electron velocity near the Fermi surface as energy independent and set $\bv = v_F\bn$, where $\bn$ is a unit vector in the direction of $\bp$.
It is possible to avoid solving the Poisson equation if one replaces $\bp$ as the argument of $f$ by $\bn$ and the energy variable $\eps = \eps_{\bp} + e\varphi(\br) - E_F$. With the new variables, the term with electric field drops out from Eq. (\ref{Boltz}), and it takes up the form 
\be
 \frac{\partial f(\bn,\eps,\br)}{\partial t}
 +
 \bv\,\frac{\partial f}{\partial\br}
 =
 \left.\hat{I}_{ee}\{f\}\right|_{\bn,\eps,\br}.
 \label{kinetic}
\ee
Physically, Eq. (\ref{kinetic}) means that in the absence of collisions, electrons near the Fermi surface just drift while retaining their total energy. 
The boundary conditions for this equation 
are $f = f_0(\eps-eV/2)$ and $f = f_0(\eps+eV/2)$ far from the orifice in the left and right half-spaces.

Equation (\ref{kinetic}) may be solved by expanding it in powers of $\alpha_{ee}$. In the zero approximation, the electrons simply propagate along straight lines while retaining their total energy $\eps$ \cite{bending}, so $f(\bn,\eps,\br)$ depends solely on whether the electron trajectory originates from orifice or not. It is convenient to use the notion of solid angle $\Omega(\br)$ at which the orifice is seen from point $\br$. In terms of this angle, the zero-approximation distribution function is
\be
 f^{(0)}_{L,R}(\eps, \bn) =
 \left\{
  \begin{array}{ll}
   f_0(\eps \mp eV/2), & \bn \notin \Omega(\br) \\
   f_0(\eps \pm eV/2), & \bn \in    \Omega(\br)
  \end{array}
 \right.
 \label{f_zero}
\ee
for the electrons in left (upper sign) and right (lower sign) half-spaces, respectively (see Fig. 1). A substitution of these expressions into Eq. (\ref{current}) results in well known expressions for the Sharvin conductance
\be
 G_{03} = \frac{\pi}{2}\, e^2\, \frac{a^2 p_F^2}{(2\pi)^2},
 \qquad
 G_{02} = \frac{e^2 p_F a}{\pi^2},
 \label{G_0}
\ee
i. e. the conductance quantum times the number of transverse  channels in the orifice.

To the first approximation in $\alpha_{ee}$, the correction to the distribution function $\delta f(\bn,\eps, \brho)$ at the orifice is readily obtained by integrating $\hat{I}_{ee}\{f^{(0)}\}$ along the trajectory of an electron with momentum direction $\bn$ that comes to point $\brho$ from infinity
\be
 \delta f(\bn,\eps,\brho)=
 \int_0^{\infty} d\tau\,
 \left.
  \hat{I}_{ee}\{\fn\}
 \right|_{\bn,\eps,\brho-v_F\tau\bn}.
 \label{df}
\ee
Here, $\tau$ is the time of motion along the trajectory. The collision integral in Eq. (\ref{df}) is nonzero only if at least one of the momenta in Eq. (\ref{I_ee}) falls within $\Omega(\br)$. As we will see below, the main contribution to (\ref{df}) comes from points $\br$ located much farther from the orifice than $a$. Hence $\Omega(\br)$ may be considered as small and the contribution to (\ref{df}) from scattering processes where more than one momentum lies in $\Omega(\br)$ may be neglected. It is well known that due to the restrictions in the momentum space, the electron--electron scattering rate exhibits singularities if the momenta of colliding electrons are either parallel or antiparallel. The latter singularity \cite{Gurzhi} is relevant to our case because it results in an anomalous behavior of the conductance. The most important contribution to Eq. (\ref{df}) arises from the collisions of electrons incident on the orifice with electrons that are injected from the other half-space and have nearly opposite momentum. Hence the integration over $\bk$ in Eq. (\ref{I_ee}) in the left and right half-spaces may be limited  to $\bk \in \Omega(\br)$. The electrons with momentum $\bk$ should be considered as injected, and the electrons with the rest of momenta $\bp$, $\bp'$, and $\bk'$, as "native" to the corresponding half-space.
 
Now we substitute Eqs. (\ref{df}) and (\ref{I_ee}) into Eq. (\ref{current}) and change the order of integrations so that the integrations over $\tau$ and $\brho$ take place first. The expression for the inelastic correction to the current assumes the form
\be
 \delta I =
 2e S_0 \alpha_{ee} \nu_d^{-2}
 \int d\eps \int d\eps' \int d\eps_1 \int d\eps_2\,
\nonumber\\
 \times
 \delta(\eps + \eps' - \eps_1 - \eps_2)\,
 F_L(\eps,\eps',\eps_1,\eps_2)\,
\nonumber\\ 
 \times
 \int\frac{d^dp}{(2\pi)^d}\,
 \delta(\eps_{\bp} - \eps - E_F)\,\Theta(v_{\perp})\,v_{\perp}
 \int\frac{d^dk}{(2\pi)^d}\,
\nonumber\\
 \times
 \delta(\eps_{\bk}-\eps'- E_F)\, 
 A(\eps_1,\eps_2, \bp + \bk)\,
 \bar{\tau}_m(\bn,\bk),
 \label{delta_I}
\ee
where the prefactor 2 appears before the integral because both half-spaces give equal contributions to $\delta I$, 
the distribution dependent factor is given by
\be 
 F_{L}(\eps,\eps',\eps_1,\eps_2)
 = 
 [1 - f_{L}(\eps)]\,
 [1 - f_{R}(\eps')]\,
 f_{L}(\eps_1)\,f_{L}(\eps_2)
\nonumber\\
  -
  f_{L}(\eps)\,f_{R}(\eps')\,[1 - f_{L}(\eps_1)]\,[1 - f_{L}(\eps_2)],\qquad
 \label{F-def}
\ee
$\Theta$ is the Heaviside step function, the quantity
\be
 A(\eps_1,\eps_2,{\bf Q})= 
 \int\frac{d^dp'}{(2\pi)^d}
 \int d^dk'\, 
 \delta(\bp' + \bk' - {\bf Q})
\nonumber\\
 \times 
 \delta(\eps_{\bp'} -  E_F  - \eps_1 )\,
 \delta(\eps_{\bk'} - E_F -\eps_2 )
 \label{A_def}
\ee
reflects the limitations in the phase space resulting from momentum conservation, and
\be 
 \bar\tau_m(\bn,\bk) =
 \frac{1}{S_0}\int d^{d-1}\rho \int_0^{\infty} d\tau\,
 \Theta[\bk \in \Omega(\brho -\tau v_F \bn)]
 \label{tau_def}
\ee
is the maximum possible time of interaction between electrons flying to the orifice along the direction $\bn$ and nonequilibrium electrons emitted from the orifice in the direction $\bk$.   
As we are interested in the linear conductance and the distribution factor $F$ vanishes at $V=0$, we have set $\varphi=0$ in all other quantities

Consider first the 3D case. The quantity $\bar\tau_m$ is easily calculated based on the geometrical considerations. In a spherical system of coordinates with the $z$ axis directed along $-\bn$, it is given by
\be
 \bar\tau_{m3}(\bn_,{\bk}) =
 \frac{8ma}{3\pi p_F}\,
 \frac
 { \cos\theta\,\cos\theta_{\bn} + \cos\phi\,\,\sin\theta\,\sin\theta_{\bn} }
 { \sin\theta \sqrt{1 - \sin^2\phi\,\sin^2\theta_{\bn}} },
 \label{tau_3D}
\ee
where $\theta$ is the zenith angle of $\bk$, $\phi$ is its azimuthal angle measured from the plane normal to the insulator, and $\theta_{\bn}$ is the angle between $-\bn$ and the normal to the insulator. Note that $\bar\tau_m$ diverges as $\theta\to 0$, so electrons moving to and from the orifice in the directions $-\bn$ and $\bn$ would be interacting during an infinitely long time. 

In 3D in the same coordinate system, the phase-space factor $A$ is given by
\be
 A_3(\eps_1,\eps_2, \bp + \bk)
 =
 (2\pi)^{-2}\,m^2\,|\bp + \bk|^{-1}
 \nonumber\\
 \times
 \Theta\!\left(|\bp + \bk| - \left|\frac{\eps_1 - \eps_2}{v_F}\right|\right)
 \approx
 \frac{m^2\,\Theta(\theta - \delta\eps_c/E_F)}
 { 2(2\pi)^5p_F\sin(\theta/2) },
 \label{A_3D}
\ee
where the quantity
$\delta\eps_c =
 (1/2)\,
 {\rm max}\!\left( |\eps_{\bp} - \eps_{\bk}|, |\eps_1 - \eps_2|\right)
$
serves as a natural cutoff for the logarithmic divergence that arises in the integral over $\theta$ when performing the integration over $\bk$ in Eq. (\ref{delta_I}).

Upon integration over $\bk$ and $\bp$ in Eq. (\ref{delta_I}),  one obtains with logarithmic accuracy
\be
 \delta I_3 =
 \frac{\alpha_{ee} m^2 a e S_0}{24 \pi^2 p_F}
 \int d\eps \int d\eps' \int d\eps_1 \int d\eps_2\,
\nonumber\\
 \times
 \delta(\eps + \eps' - \eps_1 - \eps_2)\,
 F(\eps,\eps',\eps_1,\eps_2)\,
 \ln\left(E_F/\delta\eps_c\right).
 \label{dI_3D}
\ee
The logarithm in this equation may be presented as a sum 
$
 \ln(E_F/\delta\eps_c) = 
 \ln(E_F/T) + \ln(T/\delta\eps_c).
$
As characteristic values of $\delta\eps_c$ in Eq. (\ref{dI_3D}) are of the order $\delta\eps_c \sim T$, the second logarithm in this equality may be neglected. Then the integrals over the energies are easily calculated, and one obtains the relative correction to the conductance
\be
 \frac{\delta G_3}{G_{03}} =
 \frac{\pi^2}{9}\,\alpha_{ee}\,
 \left( \frac{a}{v_F}\,\frac{T^2}{E_F} \right)\,
 \ln\left( \frac{E_F}{T} \right).
 \label{dG_3D}
\ee

In the 2D case, $\bar\tau_m$ is given by
\be
 \bar\tau_{m2}(\bn,\bk) =
 \frac{a}{v_F}\,
 \frac{\cos(\phi - \phi_{\bn})}{|\sin\phi|},
 \label{tau_2D}
\ee
where $\phi$ is the angle between $-\bn$ and $\bk$, and $\phi_{\bn}$ is the angle between $-\bn$ and the normal to the insulator. The phase-space factor $A$ now has singularities at both $\phi=0$ and $\phi=\pi$, but only the former is essential for us, so we can write it in the form 
\be
 A_2(\eps_1,\eps_2, \bp + \bk)
 =
 \frac{v_F^{-2}}{(2\pi)^2}\,
 \frac
 {\dsp \Theta\!\left( \eta \right) }
 {\dsp \cos(\phi/2)\, \sqrt{ \eta } },
 \label{A_2D}
\ee 
where $\eta = \sin^2(\phi/2) + D/4E_F^2$ and
$D = [(\eps_{\bp} - \eps_{\bk})^2 - (\eps_1 - \eps_2)^2 ]/4$. At $D >0$, the integral of $A\bar\tau_m$ over $\phi$ logarithmically diverges at $\phi=0$, so one has to introduce a cutoff $\tau_c = l_c/v_F$ in Eq. (\ref{tau_2D}). The cutoff length $l_c$ may be due, e.g., to a very weak electron--impurity scattering, which makes the electron distribution isotropic far from the orifice.

Performing the angular integration over $\phi$ and $\phi_{\bn}$ and retaining only the most singular terms, one obtains for the correction to the current
\be
 \delta I_2 =
 \frac{\alpha_{ee}}{\pi^4}\,
 ea^2m\,
 \ln\left(\frac{l_c}{a}\right)
 \int d\eps \int d\eps' \int d\eps_1 \int d\eps_2\,
\nonumber\\
 \times
 \delta(\eps + \eps' - \eps_1 - \eps_2)\,
 \Theta(D)\, F(\eps,\eps',\eps_1,\eps_2)/\sqrt{D}.
 \label{dI_2D}
\ee
The distribution-dependent factor $F$ in (\ref{dI_2D}) may be linearized with respect to $V$. With account taken of the delta function in this equation,
$$
 F(\eps,\eps',\eps_1,\eps_2) =
 \frac{eV}{T}\,
 e^{\frac{\eps+\eps'}{T}}\,
 f_0(\eps)\,f_0(\eps')\,f_0(\eps_1)\,f_0(\eps_2).
$$
Then (\ref{dI_2D}) is easily evaluated by introducing dimensionless variables $\xi_i = \eps_i/T$, and we arrive at an expression for the relative correction to the conductance
\be
 \frac{\delta G_2}{G_{02}} =
 \frac{2C_0}{\pi^2}\,\alpha_{ee}\,
 \frac{a}{v_F}\, T\,
 \ln\left(\frac{l_c}{a}\right),
 \qquad
 C_0 \approx 1.87.
 \label{dG_2D}
\ee

\begin{figure}[t]
\includegraphics[width=8.5cm]{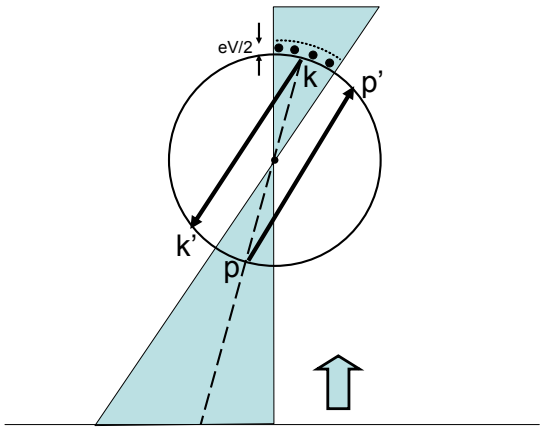}
\caption{\label{fig2} Scattering processes that result in a positive correction to the current in 2D. The gray arrow shows the direction of net particle current. Extra electrons  above the Fermi surface (black circles) are injected from the contact and reside within the  solid angle $\Omega(\br)$ (shaded area). As an extra electron in state $\bk$ collides with electron $\bp$, they are scattered into $\bk'$ and $\bp'$. Momentum conservation requires that $\bp'=-\bk'$ and $\bp=-\bk$, hence the relaxation of extra electrons decreases the number of particles that will hit the orifice.}
\end{figure}

In both 2D and 3D, the corrections to the conductance are positive, unlike the cases of electron--phonon and electron--impurity scattering. This is due to a principally different nature of electron--electron scattering, which does not result in a momentum relaxation of electrons. The positive contribution to the conductance may be considered as a sort of drag of electrons incident on the contact by electrons emitted from the other electrode with a different velocity. As shown in Fig. \ref{fig2}, the electron--electron scattering tends to smooth out the bump in the angular distribution of electrons in the right electrode while conserving their total momentum  at a given point, so the number of electrons at the opposite side of the Fermi surface decreases. In the case of a dent in the left electrode this number increases and hence the two corrections add together.

Naively, one might expect $\delta G/G_0$ to be of the order of $a/l_{ee}$, where $l_{ee}^{-1} \sim v_F T^2/E_F$ is the inverse equilibrium electron--electron scattering length. However the actual effect is much larger: $\delta G_3$ and $\delta G_2$ contain additional factors $\ln(E_F/T)$ and $E_F/T$, respectively. These factors are due to a sort of "resonance" between electrons with oppositely directed velocities moving to and from the contacts, which results from a superposition of the singularities in the scattering rate and the effective time of interaction of these electrons. The anomalously large correction to the conductance is closely related to the logarithmic singularity in the lifetime of quasiparticles in a 2D electron gas \cite{Giuliani} but is enhanced by the nonequilibrium distribution function with a strong angular dependence.
The magnitude of the effect is increased because the scattering intensity falls off away from the orifice so slowly that the actual size of the scattering region is much larger than the diameter of the orifice.

Very recently, Renard {\it et al.} \cite{Renard} observed that the conductance of wide 2D quantum contacts with a large number of transverse channels increased with temperature. The increment was proportional to the temperature and contact width. The authors attributed this increase to a quantum-mechanical interference between the scattering of electrons from impurities and from Friedel oscillations around them \cite{Zala}. We suggest that this increment may be due to purely classical effects of electron--electron scattering considered here. For the electronic density $n_s\sim 10^{-11}$ cm$^{-2}$ the gas parameter in GaAs is $r_s\sim1$ and $k_F$ is of the order of inverse screening length, hence it is reasonable to take $\alpha_{ee}\sim 1$ \cite{Altshuler}. With $a \sim\lambda_F$, we get the coefficient of $T$ in (\ref{dG_2D}) of the order of $10^{-2}K^{-1}$, which roughly agrees with \cite{Renard}. 

The authors of \cite{Renard} also report an anomalous positive magnetoresistance in low fields. This effect may be easily explained in terms of our results because the magnetic  field bends trajectories of oppositely moving electrons in opposite directions and destroys the "resonance" between them. The positive correction to the current from electron--electron interaction should be suppressed when the classical cyclotron radius $l_c=p_F/eB$ becomes of the order of the size of the interaction region $aE_F/T$.

We are grateful to V. A. Sablikov for a discussion.
This work has been supported by Russian Foundation for Basic Research, grants 06-02-72552-NCNIL-a and 07-02-00137-a.


\begin{thebibliography}{99}
\bibitem{bending}
We neglect the bending of electron trajectories because $eV\ll E_F$.
\end{thebibliography}
\end{document}